%% file: main.tex
\documentclass[sigconf]{acmart}

\AtBeginDocument{%
  \providecommand\BibTeX{{%
    \normalfont B\kern-0.5em{\scshape i\kern-0.25em b}\kern-0.8em\TeX}}}

\acmSubmissionID{123-A56-BU3}

\usepackage{amsmath,amsfonts}
\usepackage[ruled]{algorithm2e}
\usepackage{graphicx}
\usepackage{textcomp}
\usepackage{xcolor}

\usepackage{lineno,hyperref}

\modulolinenumbers[5]

\usepackage{mwe}  
\usepackage{grffile}
\usepackage{hyperref}
\usepackage{url}
\usepackage{graphicx}
\DeclareGraphicsExtensions{.eps,.pdf}
\usepackage{multirow}
\usepackage[ruled]{algorithm2e}
\usepackage{amsmath}
\usepackage{pifont}
\usepackage{enumerate}
\usepackage{tikz}
\usetikzlibrary{calc,backgrounds,arrows,trees,shapes,snakes,shadows,patterns,matrix,fit,patterns,positioning,automata,decorations.pathmorphing,decorations.pathreplacing}
\usepackage{epstopdf}
\usepackage{multicol}
\usepackage[aboveskip=-4pt,belowskip=-2pt]{subcaption}
\usepackage[font=footnotesize]{caption} 
\usepackage[ruled]{algorithm2e}
\SetKwRepeat{Do}{do}{while}

\newcommand{\mathtext}[1]{\mathrm{\textit{#1}}}

\begin{document}

%
\title{LFOC: A Lightweight Fairness-Oriented Cache Clustering Policy for Commodity Multicores}

\author{Adrian Garcia-Garcia}
\affiliation{\institution{Complutense University of Madrid}}
\email{adriagar@ucm.es}

\author{Juan Carlos Saez}
\affiliation{\institution{Complutense University of Madrid}}
\email{jcsaezal@ucm.es}

\author{Fernando Castro}
\affiliation{\institution{Complutense University of Madrid}}
\email{fcastror@ucm.es}

\author{Manuel Prieto-Matias}
\affiliation{\institution{Complutense University of Madrid}}
\email{mpmatias@ucm.es}
%
\renewcommand{\shortauthors}{Garcia-Garcia, et al.}

%
\input{abstract}

\copyrightyear{2019}
\acmYear{2019}
\setcopyright{acmcopyright}
\acmConference[ICPP 2019]{48th International Conference on Parallel Processing}{August 5--8, 2019}{Kyoto, Japan}
\acmBooktitle{48th International Conference on Parallel Processing (ICPP 2019), August 5--8, 2019, Kyoto, Japan}
\acmPrice{15.00}
\acmDOI{10.1145/3337821.3337925}
\acmISBN{978-1-4503-6295-5/19/08}


%
%


%

%

%
\maketitle

\input{introduction}

\input{background}

\input{motivation}
\input{design}
\input{experiments}

\input{conclusions}
\begin{acks}
This work has been supported by the EU (FEDER), the Spanish MINECO and CM, under grants TIN 2015-65277-R, RTI2018-093684-B-I00 and S2018/TCS-4423. Adrian Garcia-Garcia is supported by a UCM fellowship grant.
\end{acks}

%
\bibliographystyle{ACM-Reference-Format}
\bibliography{references}

\end{document}

%% file: abstract.tex
\begin{abstract}
Multicore processors constitute the main architecture choice for modern computing systems in different market segments. Despite their benefits, the contention that naturally appears when multiple applications compete for the use of shared resources among cores, such as the last-level cache (LLC), may lead to substantial performance degradation. This may have a negative impact on key system aspects such as throughput and fairness. Assigning the various applications in the workload to separate LLC partitions with possibly different sizes, has been proven effective to mitigate shared-resource contention effects. 

In this article we propose LFOC, a clustering-based cache partitioning scheme that strives to deliver fairness while providing acceptable system throughput. LFOC leverages the Intel Cache Allocation Technology (CAT), which enables the system software to divide the LLC into different partitions. To accomplish its goals, LFOC tries to mimic the behavior of the optimal cache-clustering solution, which we could approximate by means of a simulator in different scenarios. To this end, LFOC effectively identifies streaming aggressor programs and cache sensitive applications, which are then assigned to separate cache partitions. 

We implemented LFOC in the Linux kernel and evaluated it on a real system featuring an Intel Skylake processor, where we compare its effectiveness to that of two state-of-the-art policies that optimize fairness and throughput, respectively.  Our experimental analysis reveals that LFOC is able to bring a higher reduction in unfairness by leveraging a lightweight algorithm suitable for adoption in a real OS.
\end{abstract}

%% file: introduction.tex
\section{Introduction}\label{sec:intro}

Today, chip multicore processors (CMPs) constitute the dominant architecture choice for modern general-purpose computing systems and will likely continue to be dominant in the near future. Despite their benefits, these processors pose a number of challenges to the system software. One of the biggest challenges is how to mitigate the effects that come from contention on shared resources~\cite{survey-contention}. Contention occurs due to the fact that cores in a CMP are not truly independent processing units but instead typically share a last-level cache (LLC) and other memory-related resources with the remaining cores, such as a DRAM controller and a memory bus or interconnection network~\cite{ebrahimi10,camps}. Applications running simultaneously on the various cores may compete with each other for these shared resources, which could degrade their performance unevenly~\cite{selfa-pact17,camps}.

Partitioning of the shared LLC (i.e., assigning a separate cache partition with a certain size to each application in a workload) has been proven effective to mitigate shared-resource contention effects~\cite{ucp,survey-cachepart,yu-petrof,kpart,whirlpool,selfa-pact17}. Recently, cache-partitioning hardware support has been adopted in commodity Intel processors via the Intel Cache Allocation Technology (CAT)~\cite{cat}, which enables the system software to assign a certain number of cache ways to each application. Several resource management schemes that leverage this technology have been recently proposed to accomplish different objectives such as system throughput optimization~\cite{kpart}, delivering fairness~\cite{selfa-pact17}, or to improve client satisfaction in virtual environments~\cite{ginseng-atc16}. 

Our work primarily explores how to leverage Intel CAT, at the OS level, to deliver system-wide fairness, which contributes to reducing a number of  undesirable effects on the system. For example, shared-resource contention may cause an application's completion time to differ significantly across executions, depending on its co-runners in the workload~\cite{survey-contention,valencianos-tocs}.  Moreover, equal-priority applications may not experience the same performance degradation when running together relative to the performance observed when each application runs alone on the CMP~\cite{stall-time-dram,ebrahimi10}. These issues make priority-based scheduling policies ineffective~\cite{ebrahimi10}, reduce performance predictability~\cite{heechul-transc16} and may lead to wrong billings in commercial cloud-like computing services~\cite{valencianos-tocs}, where users are charged for CPU hours. Notably, unfairness also leads to uneven progress of the various threads in HPC multithreaded applications~\cite{eqp-belga,acfs-jpdc}, which seriously limits scalability.

To deliver fairness while providing acceptable system throughput, we propose LFOC, a Lightweight Fairness-Oriented Cache-clustering OS-level scheme. By using Intel CAT hardware support, LFOC dynamically creates a number of last-level-cache partitions (\textit{clusters}) based on the features of the workload, and maps applications to different clusters by catering to their degree of cache sensitivity and contentiousness.

The main contributions of our work are as follows:
\begin{itemize}
\item In order to guide the design of LFOC, we approximated --by means of a parallel simulator-- the solution to the cache-clustering problem that optimizes fairness for different workload scenarios. The exhaustive analysis of the optimal solution reveals that the key to enforce fairness lies in identifying contentious cache-insensitive (aka \textit{streaming}) applications and confine them to a reduced set of small cache partitions, so as to devote the vast majority of space in the LLC to cache-sensitive applications.  
\item Based on the insights provided by the previous analysis, we proceeded to design LFOC, which attempts to approximate the cache clustering enforced by the optimal solution. Our approach continuously monitors applications' runtime metrics with hardware performance counters and classifies applications into different categories based on cache behavior. The collected performance information is used as input to an efficient clustering algorithm. LFOC leverages a lightweight online mechanism to approximate the degree of cache sensitivity of an application that avoids costly periodic monitoring operations (i.e. measuring application performance for different cache sizes at runtime) used by other approaches~\cite{kpart}, whenever possible. 

\item We implemented LFOC in the Linux kernel and evaluated it on a real system featuring an Intel Skylake processor. In our experiments, we compare its effectiveness to that of two previously proposed policies --Dunn~\cite{selfa-pact17} and KPart\cite{kpart}--, which optimize fairness and throughput, respectively. Our analysis reveals that LFOC is able to deliver up to a 20.5\% reduction in unfairness (9\% on average) relative to Dunn (fairness-oriented clustering), and delivers higher throughput and fairness than every analyzed scheme for the vast majority of the workload scenarios considered. 

\end{itemize}

The remainder of the paper is organized as follows. Section~\ref{sec:background} presents background on cache partitioning and also discusses related work. Section~\ref{sec:motivation} presents the analysis of the optimal solution that motivates our proposal. Section~\ref{sec:design} outlines the design and inner workings of LFOC. Section~\ref{sec:experiments} covers the experimental evaluation, and Section~\ref{sec:conclusions} concludes the paper.

%% file: background.tex
\section{Background and Related Work}\label{sec:background}

In this section we first describe the metrics we considered to assess the degree of fairness and throughput provided by cache partitioning strategies. Then, we formally introduce the notion of cache partitioning and cache clustering and present some related issues. Finally, we discuss related work. 

\subsection{Metrics}\label{subsec:metrics}

To measure the performance degradation of an individual application in a multi-program workload we consider the \textit{Slowdown} metric, defined as follows:
\begin{eqnarray}
\mathtext{Slowdown}_{app}  = \frac{\mathtext{CT}_{\mathtext{part},\mathtext{app}}}{CT_{\mathtext{alone},\mathtext{app}}} \label{eq:slowdown} 
\end{eqnarray}

where \emph{$CT_{\mathtext{part},\mathtext{app}}$} denotes the completion time of application \emph{app} when it runs sharing the system under a given cache-partitioning scheme, and $CT_{\mathtext{alone},\mathtext{app}}$ is the completion time of the application when running alone on the CMP system.

The slowdown of a single-threaded application can be also defined in terms of the average number of instructions per cycle observed when it runs alone with all cache space available ($\mathtext{IPC}_{\mathtext{alone},\mathtext{app}}$) and that achieved when it runs with other applications in the workload under a certain cache-partitioning scheme ($\mathtext{IPC}_{\mathtext{part},\mathtext{app}}$):  

\begin{eqnarray}
\mathtext{Slowdown}_{app}= \mathtext{IPC}_{\mathtext{alone},\mathtext{app}}/\mathtext{IPC}_{\mathtext{part},\mathtext{app}}  \label{eq:slowdown_opc} 
\end{eqnarray}

Previous research on fairness for multicore systems~\cite{ebrahimi10,selfa-pact17} defines a scheme as fair if equal-priority applications in a workload suffer the same slowdown as a result of sharing the system. To cope with this notion of fairness, we employ the \textit{unfairness} metric, which has been extensively used in previous work~\cite{ebrahimi10,xu-sigmetrics12,pmctrack-compj,camps}. For a workload consisting of $n$ applications, this metric (lower-is-better) is defined as follows: 
\begin{eqnarray}
\textstyle \mathtext{Unfairness} = \frac{\mathtext{MAX}{(}{Slowdown}_{1}{ ,... ,Slowdown}_{n}{)}}{\mathtext{MIN}{(}{Slowdown}_{1} ,... ,{Slowdown}_{n}{)}}  \label{eq:unfairness} 
\end{eqnarray}

Notably, according to the definition of the unfairness metric, we could improve its value just by slowing down certain applications to achieve similar but potentially high slowdown figures. Clearly, this is unacceptable, as it may come at the expense of high throughput degradation, which our proposal also tries to reduce. Therefore, the value of the unfairness metric must be always reported along with system throughput figures, as we do in this article. Specifically, to quantify throughput, we used the \emph{System ThroughPut} (STP) metric~\cite{metrics-belga,selfa-pact17}, --also referred to as Weighted Speedup in~\cite{kpart}--, defined as follows:
\begin{eqnarray}
\mathtext{STP} = \sum_{i=1}^{n}\left(\frac{\mathtext{CT}_{alone,i}}{\mathtext{CT}_{part,i}}\right) = \sum_{i=1}^{n}\left(\frac{1}{\mathtext{Slowdown}_{i}}\right) \label{eq:stp}
\end{eqnarray}

\subsection{Cache partitioning vs. Cache clustering}\label{subsec:opt-cpart}

Two major strategies exist to distribute cache space among applications in a shared LLC that supports way-partitioning: \textit{cache partitioning} and \textit{cache clustering}.

\textit{Cache-partitioning} entails assigning a separate cache partition with a certain size (a specific number of ways) to each application in a workload. Let $A$ be a workload consisting of $n$ applications $\{a_{1},a_{2},\cdots{},a_{n}\}$ and let $S$ be a system that features a $k$-way last-level cache with $k\geq{}n$. A feasible cache partitioning of the LLC for $A$ on $S$ is formally defined as a set $\{w_{1},w_{2},\cdots{},w_{n}\}$ (with $\sum_{i=1}^{n}{w_{i}}=k$) where $w_{i}$ denotes the number of ways assigned to application $a_i$ ($1\leq{}w_{i}\leq{}k-n+1$).

In recent years, many cache-partitioning proposals~\cite{ucp,survey-cachepart,yu-petrof,icpp-15} have been proposed to target different optimization objectives, such as maximizing throughput, reducing energy consumption or improving fairness. These proposals are equipped with heuristic approximate algorithms specifically tailored to the target objective they pursue. In general, determining the solution to the optimal cache partitioning problem for a certain optimization objective is known to be NP-hard~\cite{survey-cachepart}, so determining the best solution via an extensive exploration of the (vast) search space is largely impractical. For example, determining the optimal cache-partitioning solution for an 8-application workload ($n=8$) on a platform with an 11-way LLC, requires the exploration of 120 solutions; the number of possible options to consider for the same workload size on a 20-way platform rises up to more than 50K.

When the number of applications exceeds the number of cache ways to share ($n>k$), partitioning the LLC is unfeasible. Two or more applications must share at least one cache way. Previous work has pointed out that even when $n\leq{}k$, the coarse granularity of partitions (in the order of MBs) available in modern processors featuring Intel CAT hardware extensions makes strict cache partitioning (no way sharing among applications) inappropriate in some cases~\cite{selfa-pact17,kpart}. Due to the finer grained cache distribution that naturally results from sharing a set of cache ways between two or more applications, we may observe better performance (and sometimes better fairness) when allowing shared ways among applications (aka. \textit{Cache Clustering}) than by means of way partitioning~\cite{icpp-15,kpart}. 

Using \textit{cache clustering}, a feasible way to assign space in the LLC to the various applications can be formally defined as follows. Let $A$ be a workload consisting of $n$ applications $\{a_{1},a_{2},\cdots{},a_{n}\}$, and let $S$ be a system that features a $k$-way shared cache. A possible distribution of LLC space on $S$ among applications in $A$ is defined as a cluster set $T=\{C_{1},C_{2},\cdots{},C_{m}\}$ and the associated set $W$ of assigned ways for each $C_{i}$ in $T$, $W=\{w_{1},w_{2},\cdots{},w_{m}\}$ where each $C_{i}$ is a disjoint subset of $A$ ($C_{i}\subseteq{}A$), subject to the following restrictions (i) $1\leq{}m\leq{}\mathrm{min}(n,k)$, (ii) $C_{1}\cup{}C_{2}\cup{}\cdots{}\cup{}C_{m}=A$, (iii) $\forall i,j, 1\leq{}i,j\leq{}m \land j>i,C_{i}\cap{}C_{j}=\emptyset$ and (iv) $\left(1\leq{}w_{j}\leq{}k-m+1\right) \land{}\sum_{i=1}^{m}{w_{i}}=k$. Intuitively, each cluster (i.e., group of applications) has a certain number of cache ways assigned to it, as indicated by the $W$ set (e.g. $C_1$ has $w_1$ ways assigned to it), which constitutes one of the possible ways to distribute the available cache ways across clusters.

As in the case of the cache partitioning, several cache clustering approaches have been recently proposed~\cite{ginseng-atc16,kpart,selfa-pact17} to pursue different optimization objectives by means of approximate algorithms. Notably, by considering the size of the search space, finding the optimal cache clustering solution for a certain objective constitutes even a harder problem to solve compared to optimal cache partitioning. Specifically, to determine the optimal solution via extensive exploration of the search space in the former case, for every possible clustering of the $A$ set --with $\mathrm{min}(n,k)$ items at the most--, we have to determine the distribution of ways among clusters that optimizes a given objective. It is worth noting that the number of possible solutions grows exponentially with $n$ and $k$. So, for example, on a system featuring a 20-way LLC the number of different cache clustering options for an 8-application workload is roughly 9M, whereas for an 11-application workload more than 5500M possible solutions exist.

\subsection{Related Work}\label{subsec:related-work}

Many researchers have attempted to mitigate the contention problem in the LLC via software and hardware techniques~\cite{survey-contention,survey-cachepart,heracles15,kpart,selfa-pact17,dirigent-asplos16,nwc-sched}. A large body of work has addressed this problem via cache-partitioning or cache-clustering approaches equipped with approximate algorithms~\cite{ucp,yu-petrof,kpart,whirlpool}. A recent survey~\cite{survey-cachepart} discusses the most effective solutions available to target various optimization objectives. 

Cache partitions can be created on systems with specific hardware support (such as Intel CAT) or by means of software-based solutions, most of which rely on page-coloring~\cite{ics-99,pact-14,xiao-eurosys09,taco16}. Page-coloring can be applied to off-the-shelf multicore platforms~\cite{palloc}, but is known to be subject to a number of limitations, which can be overcome with hardware-based cache partitioning support~\cite{selfa-pact17}. Among the different hardware alternatives, the main differences essentially lie on how to manage the number of ways for the different applications: some proposals are based on the cache replacement policy~\cite{prism-isca12,khan-hpca14,wang-micro14} while others use set sampling and duplicate cache tags to guide cache partitioning~\cite{ucp,subramanian-micro15}. In this work we propose an OS-level (also extensible to the virtual machine monitor) cache-clustering scheme that leverages hardware-based way-partitioning.

In the remainder of this section we discuss the cache-partitioning and cache-clustering policies closer to our LFOC approach.   

\subsubsection{Cache partitioning proposals}\label{subsubsec:ccp1}

UCP~\cite{ucp} is a cache-partitioning scheme that aims to improve system throughput by minimizing the total number of misses incurred by all applications in the workload on the shared last-level cache. UCP does not attempt to determine the optimal solution but instead employs an approximate algorithm referred to as \textit{lookahead}~\cite{ucp}, which uses as input the MPKI table of each application. This table stores the application's MPKI (LLC Misses Per 1K Instructions) value for any possible cache size (i.e., number of assigned cache ways under way-partitioning). In the original proposal~\cite{ucp}, UCP relies on hardware extensions to determine per-application MPKI tables at runtime. Unfortunately, more than a decade later of the original proposal, these hardware extensions have not yet been adopted in commercial platforms. Our LFOC approach relies on the \textit{lookahead} algorithm to distribute the vast majority of the space in the LLC among cache-sensitive applications, by using the per-application slowdown tables (i.e., slowdown for different number of ways relative to using the total way count) as input to the algorithm instead of MPKI tables; this enables us to provide a fairer cache distribution. The cache-partitioning algorithm  proposed by Yu and Petrov~\cite{yu-petrof} strives to reduce system bandwidth pressure. To this end, it partitions the LLC so as to minimize the total bandwidth. This algorithm relies on bandwidth consumption measurements with different cache sizes gathered offline for the various applications. As opposed to this approach, LFOC does not require offline-collected application data to function. 

\subsubsection{Cache clustering proposals}\label{subsubsec:ccp2}
More recently, different cache-clustering algorithms have been proposed~\cite{kpart,selfa-pact17} as an alternative to strict cache-partitioning. The KPart~\cite{kpart} scheme constitutes a cache clustering approach specifically designed for throughput optimization. KPart implements an iterative algorithm that creates and merges application clusters via hierarchical clustering. To decide which clusters must be merged in each iteration of the loop, and how to distribute the available ways among clusters (inter-cluster way-partitioning), the scheme leverages the distance metric proposed in~\cite{whirlpool} as well as UCP's \textit{lookahead} algorithm~\cite{ucp}. The application of \textit{lookahead} and the evaluation of the distance metric relies on the ability to determine MPKI tables and IPC tables (i.e. number of Instructions Per Cycle for different cache sizes) online for each application. As explained in Sec.~\ref{sec:design}, LFOC requires to gather a smaller amount of performance information than KPart, and avoids to perform costly cache way sweeps periodically, thus effectively reducing the overhead. 

In~\cite{selfa-pact17}, Selfa et al. propose the Dunn cache-partitioning policy, designed to improve fairness. This approach groups applications into clusters by applying the \textit{k-means} clustering algorithm, 
using the fraction of core stalls caused by L2 cache misses incurred by the applications as the metric to guide clustering. (In our experimental platform this information can be obtained with the \texttt{STALLS\_L2\_MISS} performance counter event.) We should highlight that, according to the definition provided in Sec.~\ref{subsec:opt-cpart}, this strategy does not strictly  constitute a pure cache-clustering approach, since the cache partitions it creates may overlap with each other. This overlapping can create unpredictable interactions between applications that belong to different clusters~\cite{kpart}. 

In our experimental evaluation, we compare the effectiveness of our proposal (LFOC) to the Dunn and KPart approaches, and demonstrate that LFOC delivers higher reductions in unfairness than these approaches across the board. We should highlight that Dunn and KPart are user-level clustering approaches, unlike LFOC, which was implemented in the OS kernel. User-level solutions may incur higher overheads since they make extensive use of system calls to access privileged resources such as performance monitoring hardware and cache partitioning facilities, which are handled by the OS. LFOC, by contrast, accesses these facilities directly via a lightweight kernel-level API. Moreover, because using floating-point (FP) is problematic at the kernel level~\cite{love-lkd}, our implementation of LFOC is free of any FP operation, as opposed to KPart's~\cite{kpart-github}, which heavily relies on it.

%% file: motivation.tex
\section{Analysis of the optimal cache-clustering solution}\label{sec:motivation}

As stated earlier, the design of our approach is inspired by the behavior of the optimal cache-clustering solution that optimizes fairness. In this section we provide an analysis on the optimal solution, which we could approximate for different workload scenarios by using the PBBCache simulator~\cite{pbbcache-github}. This simulation tool relies on offline-collected application performance data obtained for different cache sizes (e.g., instructions per cycle, memory bandwidth consumption, etc.) to determine the degree of throughput, fairness and other relevant metrics for a workload under a particular partitioning algorithm on a given platform. A key feature of the PBBCache simulator is its ability to determine the optimal cache-partitioning and optimal cache-clustering solutions for different optimization objectives (e.g. throughput or fairness) using a parallel branch-and-bound algorithm. To obtain the application slowdown, which is necessary to determine the degree of fairness, PBBCache accounts for the performance degradation due to both cache sharing and memory-bandwidth contention (to this end it uses a variant of the probabilistic model proposed in~\cite{morad-jpdc16}). 

To carry out our analysis with the simulator, we used performance counters to gather the average value of different runtime metrics with varying cache sizes for applications from the SPEC CPU2006 and CPU2017 suites running alone on a real system featuring an Intel Skylake processor with an 11-way 27.5MB LLC. (More information on this platform can be found in Sec.~\ref{sec:experiments}.) The offline-collected metric values, which correspond to the execution of the first 1500 billion instructions of the aforementioned benchmarks, are used as input to the simulator. This information is used to determine the optimal clustering solution for fairness, namely, the solution to the optimal cache-clustering problem that obtains the optimal (minimal) unfairness value for the maximum throughput (STP) attainable. 

\begin{figure}[tbp]
\centering
\includegraphics[width=0.32\textwidth]{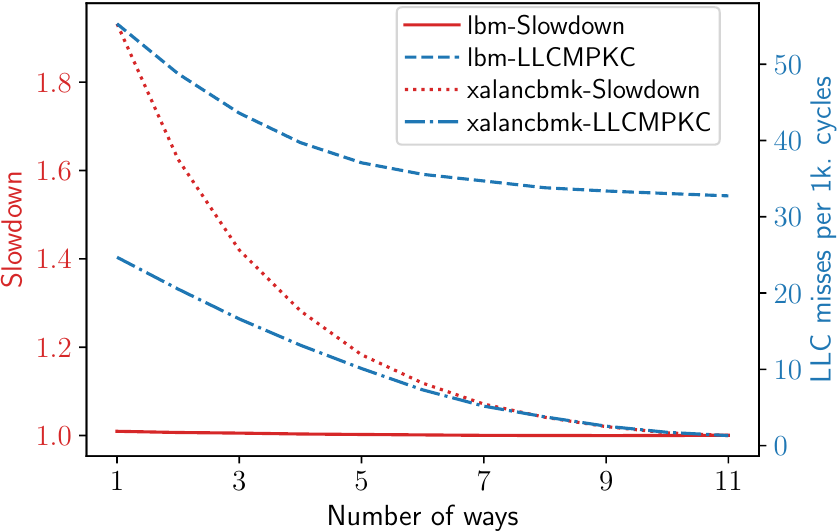}
\vspace{-0.2cm}
\caption{Slowdown and LLCMPKC for different way counts\label{fig:osci}}
\end{figure}

For our experiments we considered randomly-generated multiprogram workloads including different number of SPEC CPU applications (from 4 to 16). According to the performance data collected offline we classify applications into three classes based on their degree of cache sensitivity and contentiousness: \emph{Cache-sensitive}, \emph{light sharing} and \emph{streaming} programs. At a high level, the cache-sensitive category is used for those programs that experience high performance drops as we reduce the number of cache ways allotted to them; this is not the case for \emph{light sharing} and \textit{streaming} applications. Streaming programs are characterized by exhibiting a low slowdown for almost all way allocations, while incurring a high number of LLC misses per cycle. Applications of this kind are cache insensitive, and typically act as aggressor programs to cache-sensitive applications co-located on the same cache cluster, as the performance of the latter can be degraded substantially. Light-sharing programs are neither cache sensitive nor aggressive to others (the working set typically fits in the per-core private cache levels). Table~\ref{tab:classif} summarizes the criteria we followed to make this classification on our experimental platform, which is based on two offline-collected metrics: the application slowdown --relative performance with respect to using the entire LLC space--  and the number of LLC Misses Per Kilo Cycles (LLCMPKC). As an example that illustrates the differences in the behavior of a streaming application (\texttt{lbm}) and that of a cache-sensitive one (\texttt{xalancbmk}), Fig~.\ref{fig:osci} shows how the slowdown and the LLCMPKC varies with the amount of ways allocated to each application.

\begin{table}
\small
  \caption{Classification of applications based on cache behavior}
  \label{tab:classif}
  \begin{tabular}{cl}
    \toprule
    \bf Type & \bf Criterion\\
    \midrule
    Streaming & ($Slowdown\leq{}1.03$ and $LLCMPKC\geq{}10$)\\  
              & in at least one way assignment, and\\
              & $Slowdown < 1.06$ in all way assignments\\
   Sensitive & If not streaming and $Slowdown\geq{}1.05$ \\ 
					& for a number of ways $\geq 2$ \\
	Light sharing & Not streaming and not sensitive\\
  \bottomrule
\end{tabular}
\end{table}

\begin{figure}[tbp]
\centering
\includegraphics[width=0.34\textwidth]{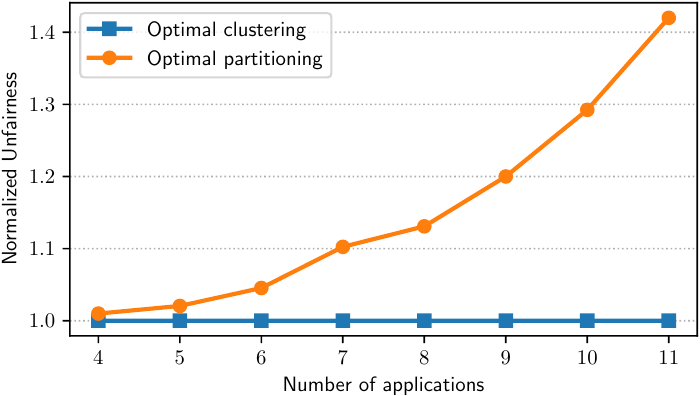}
\vspace{-0.15cm}
\caption{Comparison of optimal clustering vs optimal partitioning. \label{fig:clus_vs_part}}
\end{figure}

After a thorough analysis of the optimal cache-clustering and optimal cache-partitioning solutions provided by the simulator for the various workloads, we draw the following major insights:

\begin{itemize}
\item In most cases, the cache-clustering solution that optimizes fairness isolates all streaming applications in a reduced set of ways (no greater than 2 in any workload). In many scenarios, a single 1-way cluster is used to confine all streaming programs.
\item This same solution maps light-sharing programs onto different clusters following a hardly predictable pattern. More importantly, by conducting additional analyses with the simulator, we observed that moving individual light-sharing applications to different clusters has very little impact on throughput and fairness.      
\item As expected, by catering to the definition of the unfairness metric, the amount of ways assigned to cache-sensitive applications is critical for both throughput and fairness. Recall that the unfairness metric factors in the maximum slowdown observed across applications in the workload, and sensitive benchmarks typically experience a very high performance degradation if their cache size requirements are not fulfilled. 
\item The benefit that comes from assigning separate cache partitions (even optimally) to individual applications decreases dramatically as the number of applications gets closer to the number of cache ways. As an illustrative example, Fig.~\ref{fig:clus_vs_part} shows the average unfairness delivered by the optimal partitioning solution, normalized to that of the optimal clustering solution for different workload sizes. As observed, optimal cache-partitioning suffers from increased unfairness as the workload size grows. When the application count matches the number of ways, each application can be assigned only one way under strict cache-partitioning -this is the only feasible option-, which gives rise to high unfairness in most workloads. Our overarching conclusion is that cache-clustering policies are clearly superior to cache-partitioning approaches as the ratio of the number of ways to the number of applications decreases.  
\end{itemize}

\begin{figure}[tbp]
\centering
\includegraphics[width=0.36\textwidth]{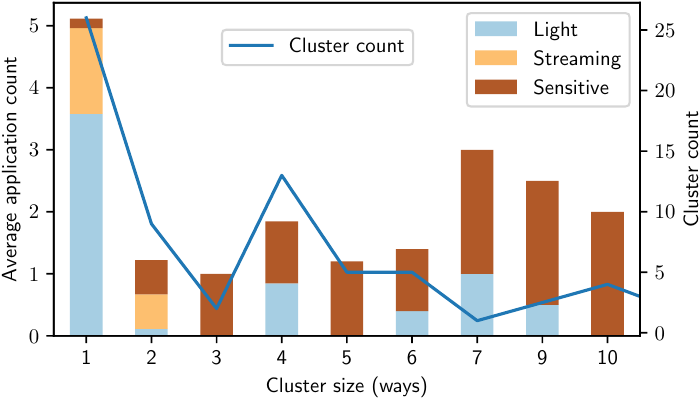}
\vspace{-0.15cm}
\caption{Cluster count and breakdown of applications into the different categories for each cluster size.\label{fig:break}}
\end{figure}

To further illustrate the general behavior of the optimal clustering solution, Fig.~\ref{fig:break} reports the average application count per cluster size, as well as the total number of clusters --grouped by its size (in ways)-- that the solution builds for a subset of the workloads we explored: 20 randomly selected program mixes made up of 10 applications each. The data reported in the figure confirms the first three aforementioned observations. First, streaming applications are typically confined in clusters with just one way allocated to them. In relative numbers, more than 87\% of streaming application instances are assigned to this kind of clusters, while the remaining ones are allocated to 2-way clusters. Second, we can observe that light sharing applications are mapped to clusters with very different size; however, the vast majority of these programs are mapped to 1-way clusters. Third, the results reveal that cache-sensitive applications are predominantly present in big cache clusters. Specifically, more than 77\% of the sensitive application instances are assigned to clusters with 4 or more ways. Finally, as is evident, 1-way clusters with a high number of applications are often present in the optimal solution for the various workloads. 

%% file: design.tex
\section{Design and Implementation}\label{sec:design}

In this section we begin by describing how our clustering algorithm works at a high level. Then we proceed to indicate how applications are classified at runtime by leveraging data from hardware performance monitoring counters.

\begin{algorithm}[t]
\fontsize{9}{9}\selectfont
\KwIn{$ST$, $CS$, and $LS$ represent the sets of streaming, cache-sensitive and light-sharing applications, respectively; $max\_streaming\_way$ and $gaps\_per\_streaming$ are configurable parameters of LFOC (default value 5 and 3, respectively), $nr\_ways$ is the number of ways of the LLC.}
\BlankLine
\SetKwFunction{funname}{LFOC\_partitioning}{
\SetKwFunction{bbound}{kk}
\SetKwFunction{look}{lookahead}
\textbf{function} \funname{$ST$,$CS$,$LS$,$nr\_ways$}:

	\If{$|CS| == 0$}{
		Create a single cluster $S$ consisting of $nr\_ways$\;
	 	Map all applications in $ST \cup LS$ to $S$\;
	 	\Return{$\{S\}$}
	}		
	$Clusters \leftarrow{} \varnothing$\;
	$ways\_for\_streaming \leftarrow{} \mathrm{\textit{min}}(2,|ST|/max\_streaming\_way)$\;
	$r=\lceil|ST|/ways\_for\_streaming\rceil$\;
	\For{$i \gets 1$ \textbf{to} $ways\_for\_streaming$}{
		Add a new 1-way cluster $C$ to $Clusters$\;
		Map up to $r$ apps from $ST$ to $C$\;
		Remove assigned apps from $ST$\;
	}	
	\{ Use slowdown tables of CS apps. as input to lookahead \}
	$W\leftarrow{}$\look{$CS,nr\_ways-ways\_for\_streaming$}\;
	\For{$i \gets 1$ \textbf{to} $|CS|$}{
		Add a new cluster $C$ with $W[i]$ ways to $Clusters$\;
		Map application $i$ in $CS$ to $C$\;
	}
	$idx \leftarrow{} 0$\;
	\While{$|LS|>0$ \textbf{and} $idx < ways\_for\_streaming$}{
		$TargetC \leftarrow{} Clusters[idx]$ \;
		$gaps\_available \leftarrow{} r - |TargetC|*gaps\_per\_streaming$\;
		 \If{$gaps\_available > 0$}{
		 	Map up to $gaps\_available$ apps from $LS$ to $TargetC$\;
		 	Remove assigned apps from $LS$\;
		 }	
	}

	Distribute remaining applications in $LS$ in a round-robin fashion among non-streaming clusters\;  
	\Return{$Clusters$}
}
\caption{Cache-clustering algorithm used by LFOC \label{alg:clustering}   }
\end{algorithm}

\subsection{Algorithm outline}

LFOC has been implemented on Linux as an extension of the OS scheduler. Specifically, it has been bundled in a loadable kernel module as a \textit{monitoring plugin} of the PMCTrack tool~\cite{pmctrack-compj}. This tool provides a kernel-level API to access privileged hardware facilities such as performance monitoring counters (PMCs) and Intel CAT features (i.e., HW support for cache way partitioning). 

LFOC classifies applications at runtime into three classes based on its cache behavior --\textit{light sharing}, \textit{streaming} and \textit{sensitive}-- and assigns each application to a certain cache partition whose size is determined dynamically based on the properties of the workload. 

When an application enters the system its cache behavior is unknown. To this end, a special \textit{unknown} class is assigned to the application right after being spawned. At the beginning of the execution, each thread has to go through a warm-up period (3 sampling intervals in our experimental setting).  Any performance information gathered with hardware counters during the warm-up period is not used to classify applications, so as to mitigate mispredictions associated with cold-start effects (e.g. the number of cache misses typically spikes intermittently at the beginning of the execution). 

Periodically, our scheduler extension activates the partitioning scheme depicted in Algorithm~\ref{alg:clustering}, which relies on the conclusions of the analysis presented in Sec.~\ref{sec:motivation}. Overall, the algorithm reserves up to two cache ways to map streaming programs. The remaining cache ways are distributed among cache sensitive applications, which are then assigned to separate cache partitions. The size of these partitions is determined by means of the lookahead algorithm~\cite{ucp}, using as input the slowdown curve for each application (i.e., slowdown registered for different cache ways) built by using IPC values obtained online. With this cache-way distribution for cache-sensitive applications, LFOC attempts to fulfill their cache requirements based on the degree of cache sensitivity. Finally, light sharing applications are distributed among the various partitions, by attempting to populate partitions with streaming applications first, as the optimal solution typically does.

\subsection{Application Classification}

Once the warm-up period has elapsed for a particular application, LFOC enters a \textit{sampling mode} whose goal is to determine the application class based on its performance sensitivity to the amount of space assigned in the LLC. This is crucial to decide on the share of the total cache space to be alloted to the application, as well as to determine what co-runners in the workload (if any) must be assigned to the same cache partition~\cite{survey-cachepart}. 

The sampling mode is inspired by the technique proposed in~\cite{kpart}, which operates as follows. Two non-overlapping complementary cache partitions covering the entire LLC space are created; the first one, referred to as the \textit{sampling partition}, is reserved for the application that triggered the transition into sampling mode, and the other one is devoted to the remaining applications. To determine the application class --based on the classification criteria presented in Sec.~\ref{sec:motivation}-- the value of various hardware events (i.e., number of instructions retired, cycles and LLC misses) is gathered with PMCs as we vary the size of the sampling partition. Notably, for sensitive applications we also obtain the slowdown curve, which is required to create partitions for these applications, as depicted in Algorithm 1. Once the sampling process terminates, LFOC transitions back into the normal operating mode described earlier.

In the original approach~\cite{kpart}, the size of the first partition is varied from the number of ways minus one to 1, whereas the size of the other partition (complementary) increases accordingly. This full sweep is required by the dynamic version of the KPart clustering approach~\cite{kpart}, which relies on the ability to accurately determine the IPC and the number of LLC Misses Per Kilo Instructions (LLCMPKI) for each way count and for \textit{every application} in the workload. We observed that this approach introduces substantial overheads due to the fact that the cache assignment enforced during the sampling mode is typically suboptimal. The sampling application receives a progressively smaller amount of cache space, while the remaining applications share a increasingly bigger cluster. This usually leads to performance/fairness degradation especially when cache sensitive applications and streaming programs are included in the workload.  

To overcome these shortcomings, LFOC immediately puts a stop to the sampling process --performed in the opposite direction (i.e. the size of sampling partition increases gradually rather than decreasing)-- in scenarios where varying the size of the sampling partition further provides no useful information to the clustering algorithm. Firstly, when the LLC miss rate falls below a certain low threshold, performance does not increase much when allotting more cache space to the application, so we expect IPC values --used to construct slowdown tables-- to remain very close beyond that point. Secondly, streaming applications typically exhibit a very low increase in performance when granting more cache space to them. In these scenarios, LFOC interrupts the sampling process and proceeds to determine the application class. In practice, to successfully identify many streaming and light sharing applications --whose slowdown curves are not needed by LFOC-- only a few way counts must be explored. When the sampling process is canceled (due to the first criterion) for a sensitive application, LFOC uses the last IPC sample gathered to approximate the performance with higher way counts, which is necessary to build the entire slowdown table.

\begin{figure}[tbp]
\centering
\includegraphics[width=0.47\textwidth]{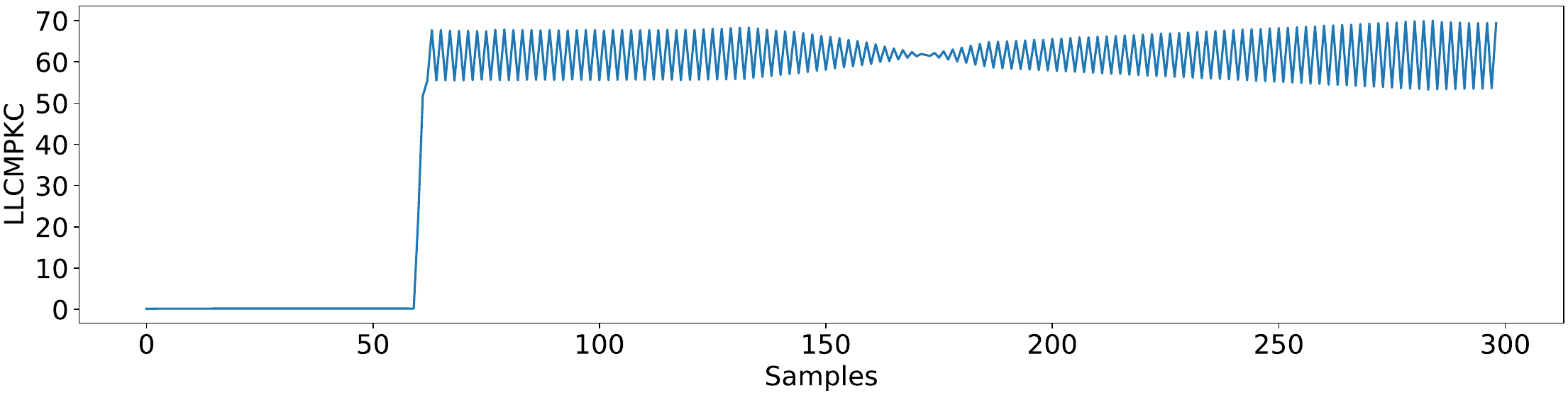}
\vspace{-0.2cm}
\caption{LLCMPKC captured at the beginning of the execution of \texttt{fotonik3d}.\label{fig:fotonik3d}}
\end{figure}

Because an application may exhibit different program phases at runtime, the initial classification may not be representative throughout the execution. For example, Fig.~\ref{fig:fotonik3d} shows the LLCMPKC of the streaming \texttt{fotonik3d} application over time, where a short light-sharing phase precedes the streaming behavior that the program exhibits for the vast majority of the execution. Failing to determine application classes accurately could lead to suboptimal cache-partitioning for certain time periods, and hence to unfairness. Triggering the sampling mode periodically helps to mitigate this issue, but, unfortunately, it backfires by introducing substantial overheads. To determine application classes at runtime in a lightweight manner, LFOC triggers a transition into the sampling mode only in the event that the application class has likely changed. To this end, the OS continuously monitors for each application the value of the LLCMPKC metric and the fraction of pipeline stall cycles incurred due to long-latency memory accesses (approximated via the \texttt{STALLS\_L2\_MISS} performance counter event, also used in~\cite{selfa-pact17}), and leverages a few heuristics to capture class changes. In particular, a class change is signaled for a light sharing application if it enters a memory-intensive phase, namely,  the average LLCMPKC measured over the last five monitoring periods exceeds a \texttt{high\_threshold} (10 in our experimental setting, as reported in Table~\ref{tab:classif} for streaming-like behavior) or the average fraction of long-latency memory-access stalls is greater than 25\%.  This approach filters out spikes in the aforementioned metrics while effectively identifies memory-intensive phases. Conversely, for streaming programs, which LFOC typically assigns to cache clusters consisting of one way, the sampling mode is engaged if its average LLCMPKC falls below a  \texttt{low\_threshold} (defined as 30\% of \texttt{high\_threshold}). Finally, for sensitive applications, LFOC associates a \textit{critical size}, defined as the amount of cache space where the slowdown falls below 5\%. The \textit{critical size} is determined during the last sampling period triggered by the application. Essentially, a class change is signaled for sensitive applications when they enter a stable non-memory intensive phase (inverse of the criterion presented earlier for light-sharing applications) for effective cache allocations\footnote{The amount of cache space used by an application is gathered by leveraging the Intel Cache Monitoring Technology.} smaller than the critical size, or when the  average LLCMPKC is higher than \texttt{high\_threshold}  for an amount of cache space bigger than the critical size.

%% file: experiments.tex
\section{Experiments}\label{sec:experiments}

To assess the effectiveness of our OS-level cache-clustering approach we implemented it in the Linux kernel v4.9.160. For the experiments we used a server platform featuring a Xeon Gold 6138 ``Skylake'' multicore processor where cores run at 2Ghz. This processor integrates an 11-way 27.5MB last level (L3) cache that supports way-partitioning; each core features a 64KB L1 cache and a 1MB L2 cache (private levels).


On this platform we carried out a experimental comparison of LFOC with the stock Linux kernel --it does not partition the LLC--, and with the Dunn~\cite{selfa-pact17} and KPart~\cite{kpart} cache-partitioning policies, specifically designed to optimize fairness and system throughput respectively. To perform a fair comparison with previous approaches we used a similar methodology as that described in the corresponding articles~\cite{selfa-pact17,kpart}. Essentially, we conduct experiments with HPC multiprogram workloads consisting of a mix of single-threaded benchmarks from SPEC CPU, and run each program for a fixed number of instructions (150 billion instructions in our setting). Specifically, we ensure that all applications in the mix are started simultaneously, and when one of them completes the corresponding instructions, the program is restarted repeatedly until the longest application in the set completes three times. We then measure unfairness and STP (throughput), by using the geometric mean of the completion times for each program. 

\begin{figure*}[tbp]
\centering
\vspace{-0.3cm}
\ifclasses
\includegraphics[width=0.87\textwidth]{all_classes-makeup-rev.pdf}
\else
\includegraphics[width=0.87\textwidth]{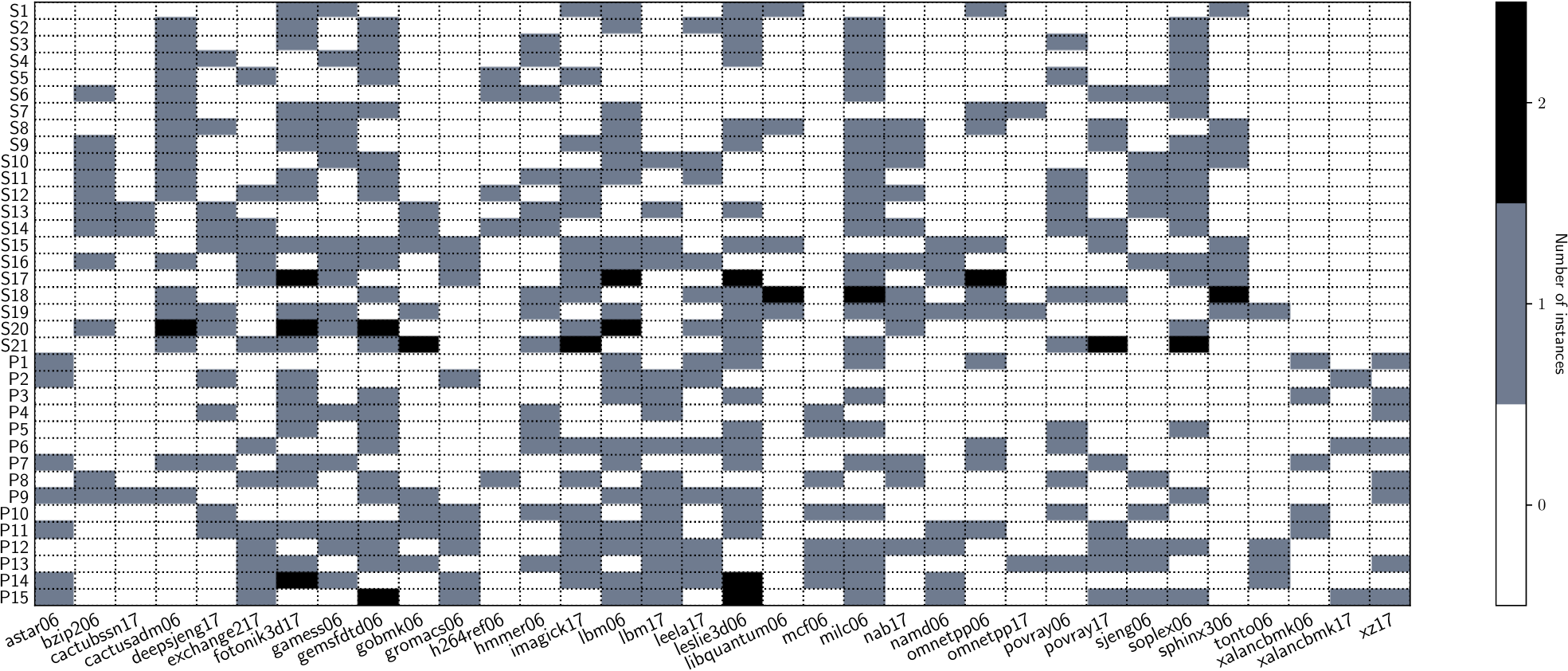}
\fi
\caption{Multiprogram workloads used for our experiments. Each matrix cell indicates the number of instances of a benchmark (x-axis) in a workload  (y-axis).\label{fig:composition}. %
\ifclasses
The number in parentheses by each benchmark's name indicates the overall application's class: light sharing (0), streaming (1) or sensitive(2). %
\fi
}
\end{figure*}

Fig.~\ref{fig:composition} depicts the composition of the 36 randomly generated workloads we used in our experiments, which are made of benchmarks from the SPEC CPU2006 and CPU2017 suites. Note that we selected applications from both suites to experiment with a wider range of streaming and cache-sensitive programs, as most benchmarks in both suites exhibit a light sharing, cache-insensitive execution profile on our platform. This is caused in part due to the coarse granularity of the cache partitions we can create on this system: the smallest partition is as big as 2.5MB. As is evident, we considered workloads of 8, 12 and 16 applications each, so as to analyze the impact that the workload size has on the fairness improvement achieved by each approach.

In this section we first evaluate the effectiveness of the cache-clustering algorithms associated with the KPart, Dunn and LFOC approaches. We then proceed to analyze how well dynamic clustering approaches deal with the time-changing behavior of the applications in different workloads.

\newif\ifclasses
\classesfalse

\def\tamfig{0.82}

\begin{figure*}[tbp]
\centering
\includegraphics[width=\tamfig\textwidth]{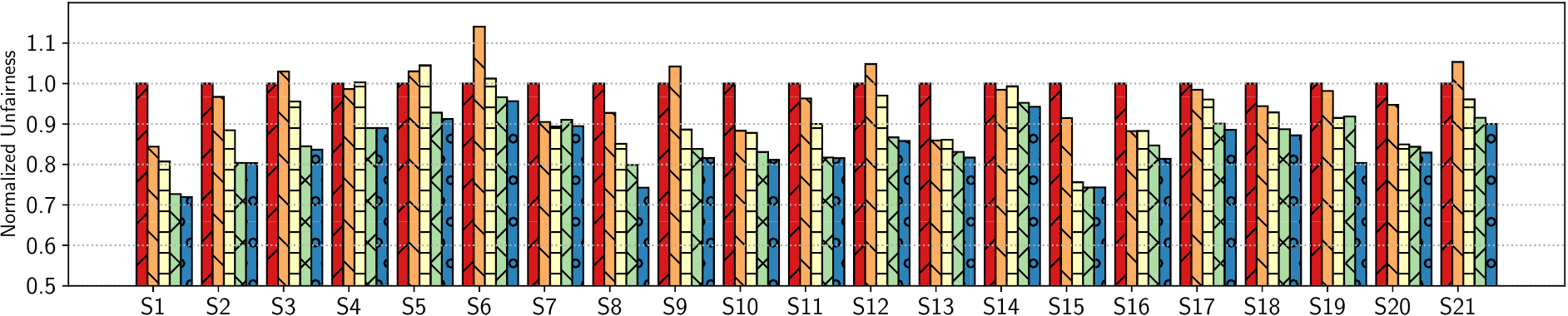}
\includegraphics[width=\tamfig\textwidth]{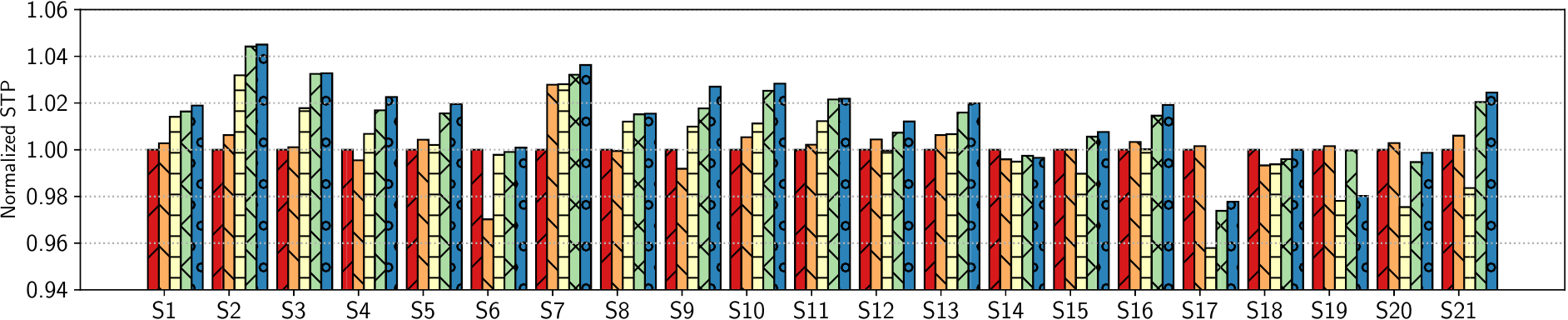}
\includegraphics[width=0.4\textwidth]{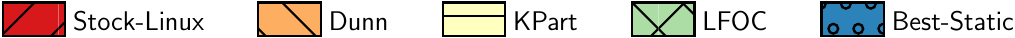}
\caption{Normalized unfairness and STP values obtained by the static version of the various clustering algorithms.\label{fig:static}}
\end{figure*}

\begin{figure*}[tbp]
\centering
\includegraphics[width=\tamfig\textwidth]{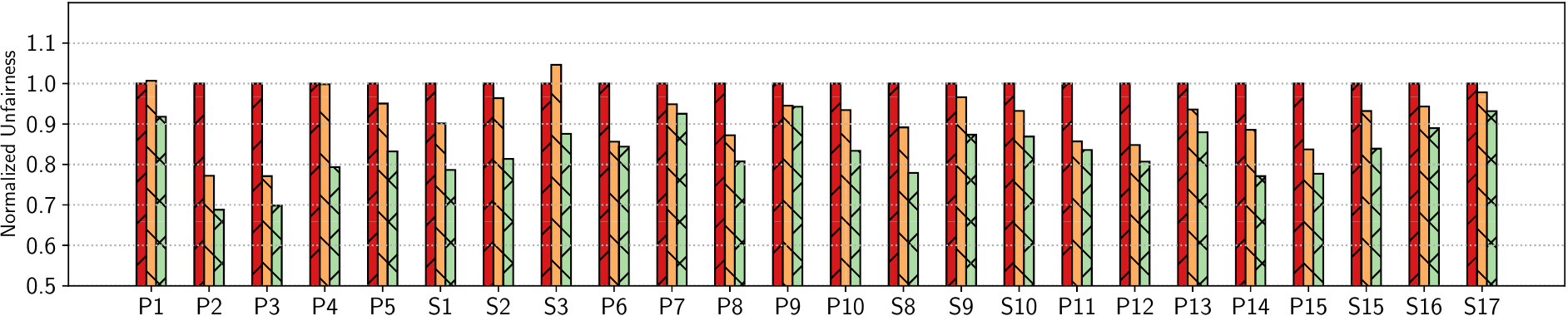}
\includegraphics[width=\tamfig\textwidth]{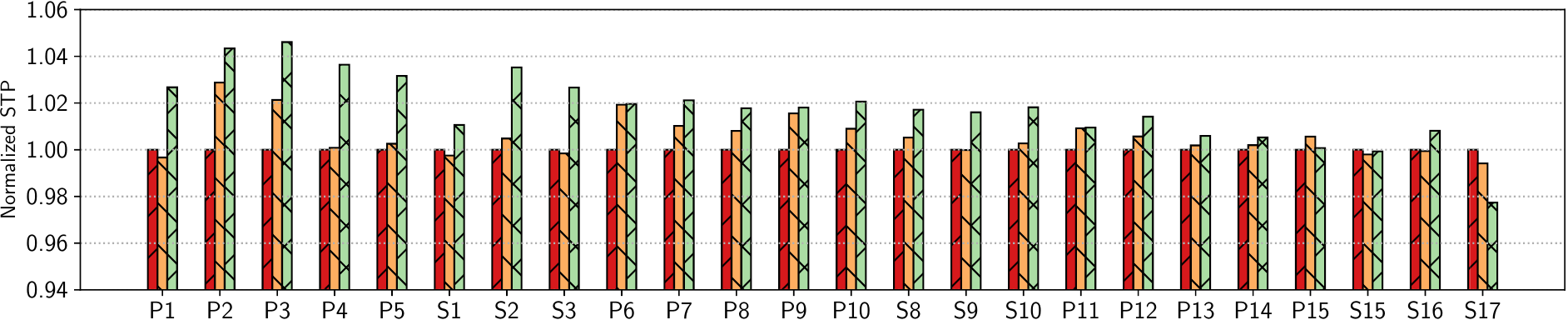}
\includegraphics[width=0.25\textwidth]{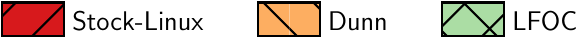}
\caption{Normalized unfairness and STP values delivered by the dynamic cache clustering approaches\label{fig:dynamic}}

\end{figure*}

\subsection{Evaluation of Clustering Algorithms}

Our goal is to measure the degree of fairness and throughput delivered by a certain clustering strategy alone (i.e. how applications are grouped into shared or separate clusters according to their runtime properties) putting aside the associated overheads due to algorithm execution, performance monitoring and cache allocation. We do account for these overheads in the experiments of the next section. 

To assess the effectiveness of each clustering algorithm, we consider workloads consisting of applications whose behavior falls in a clear class (cache sensitive, streaming or light sharing) for the vast majority of the execution (S$i$ workloads in Fig.~\ref{fig:composition}). For the analysis, we implemented the clustering algorithms used by KPart, Dunn and LFOC on top of the simulator described in Sec. ~\ref{sec:motivation}, which accepts as input the average value of various performance metrics gathered offline for different cache sizes.  To conduct the corresponding experiments, we launch the simulator prior to the execution of each workload to retrieve the cache-partitions and application-to-partition mappings imposed by a certain clustering strategy. Then, we enforce the corresponding cache partitions on a per-process manner from user-space, using the PMCTrack tool~\cite{pmctrack-compj}, and proceed to launch the workload, which will use the same \textit{static} cache configuration throughout the execution. For comparison purposes, we have also gathered the results of an ideal cache-clustering policy, referred to as \text{Best-Static}, which establishes the cache-partitions and application-to-cluster mappings based on the optimal fairness solution determined by the simulator.

Fig.~\ref{fig:static} shows the degree of unfairness and throughput delivered by the different clustering strategies; the values have been normalized to the results of Stock-Linux (no cache partitioning). The results reveal that the Dunn approach, designed to optimize fairness, exhibits a non-uniform behavior across workloads; for some program mixes it is capable to reduce unfairness up to 15.5\% , but for others it causes substantial fairness degradation (by a factor of up to 1.14x) relative to Stock-Linux. We found that this is due to its exclusive reliance on the \texttt{STALLS\_L2\_MISS} performance event; the higher the value of this event, the higher the number of cache ways allotted by Dunn to the application ~\cite{selfa-pact17}. More specifically, we observed that some streaming (aggressor) cache-insensitive applications, such as \texttt{GemsFDTD} or \texttt{fotonik3d} exhibit high values of this event, as their performance is greatly affected by memory accesses. These applications can be mapped together to the same (or overlapping) cache partitions with highly sensitive programs, such as \texttt{soplex} or \texttt{omnetpp}, for which the event reaches similar figures, leading to performance degradation and unfairness. Based on this insight, we conclude that using the \texttt{STALLS\_L2\_MISS} event alone is not enough to drive fairness-aware partitioning policies.

We also observe that KPart's clustering algorithm, designed to optimize throughput, brings modest throughput gains\footnote{In the original paper, the authors report a 24\% average increase in throughput on a different platform, with workloads whose composition was not disclosed. } in the workloads we explored (up to 3\%). However, this approach does bring substantial unfairness reductions (8.6\% on average). Nonetheless, we observe that LFOC's simpler and more lightweight partitioning algorithm provide substantially better fairness than KPart for the vast majority of the workloads (up to 27.3\%, and 14\% on average relative to Stock Linux). At the same time, LFOC achieves higher throughput than KPart across the board, and performs in a close range (1.8\% on average) of the \textit{Best Static} approach (our approximation to the optimal policy in these workload scenarios).

\subsection{Study of the dynamic policies}

For the evaluation in this section we used our OS-level implementation of LFOC, and also created a user-level implementation of Dunn, as it was originally proposed~\cite{selfa-pact17} as a user-level cache clustering approach. A good property of Dunn is the fact that it only requires the continuous monitorization of the \texttt{STALLS\_L2\_MISS} performance event for the various applications over time. The simplicity of the Dunn approach stands in contrast to the higher complexity of KPart, which relies on the ability to accurately gather a substantial amount of performance information online for each application (LLCMPKI and IPC values for every possible cache-way count) in order to apply the clustering algorithm.

\begin{table}[h]
	\begin{center}
	\small\addtolength{\tabcolsep}{-2pt}
\caption{Average execution time (in ms) of the KPart and LFOC algorithms\label{table:kpart}}
	\scalebox{0.9}{%
	\begin{tabular}{|c|c|c|c|c|c|c|c|c|}
	\hline
	    \#Apps. & 4 & 5 & 6 & 7 & 8 & 9 & 10 & 11 \\ \hline
	    LFOC & 0.00151 & 0.00154 & 0.00163 & 0.00174 & 0.00174 & 0.00182 & 0.00191 & 0.00216 \\ \hline
	    KPart & 0.51800 & 0.79600 & 1.21800 & 1.48100 & 2.01200 & 2.74200 & 3.32000 & 4.14000 \\ \hline
	\end{tabular}}
	\end{center}
\end{table}

In an attempt to evaluate the dynamic version of KPart --referred to as KPart-Dynaway~\cite{kpart}-- we considered the user-level implementation created by the authors~\cite{kpart-github}. Unfortunately, this implementation, which consists of roughly 4K lines of C++ code and makes intensive use of the Armadillo linear algebra library, was specifically tailored to the hardware platform where the authors conducted the experiments~\cite{kpart}, and makes numerous assumptions that do not apply to our experimental setting (e.g., the number of cache ways should be no smaller than the number of applications in the workload). Due to these platform-specific assumptions and other issues --as yet unidentified--, the execution of KPart-Dynaway crashes shortly after the partitioning algorithm is executed for the first time, preventing us from launching any of the workloads we considered for the evaluation. We leave for future work the adaptation of this somewhat complex implementation for our platform, and its complete evaluation. Nevertheless, to highlight the enormous difference between the complexity of KPart's partitioning algorithm and the one used by our approach for different number of applications, Table~\ref{table:kpart} shows the execution time for both algorithms (compiled with aggressive optimizations) for different workload sizes. We were able to gather that information from KPart's implementation by instrumenting the code of the partitioning algorithm that completes successfully (for workloads with less than twelve applications) right before the program's crash. As is evident LFOC's execution time (2$\mu$s) is up to three orders of magnitude smaller than KPart's, which can take over 4ms to complete for 11 applications (slightly longer than the default timer tick in the Linux kernel). As we showed in the previous section, this increased complexity does not enable KPart to provide better fairness than our lightweight approach. 

In our OS-level implementation of LFOC we sample performance counters every 100M instructions during the normal operation mode (see Sec.~\ref{sec:design}) and every 10M instructions during the sampling mode. Using a shorter instruction window for the sampling mode makes it possible to reduce the time required to complete sampling.  Notably, we observed that in most cases full cache-way sweeps are not required during LFOC's sampling mode as the partitioning algorithm (as explained in Sec.~\ref{sec:design}) does not require detailed per-way metrics for all applications, as opposed to KPart. In our experiments, the partitioning algorithm for both Dunn and LFOC is executed every 500ms, as this is the setting used in the original evaluation of the Dunn approach~\cite{selfa-pact17}. 

Fig.~\ref{fig:dynamic} shows the normalized unfairness and throughput values delivered by the Dunn and LFOC (dynamic) approaches for different workloads. Note that in this case, we considered additional program mixes (P$i$ workloads) that include applications such as \texttt{xz}, \texttt{astar}, \texttt{mcf} or \texttt{xalancbmk}, which exhibit distinct long-term program phases with varying degree of memory intensity. Some of these applications go through highly cache sensitive phases, so Stock-Linux delivers higher unfairness values in these scenarios. That is the reason why  Dunn exhibits a slightly fairer behavior under these circumstances relative to the scenario considered in the previous section. Yet, LFOC is capable to provide better throughput than Dunn, and improves fairness over Dunn across the board (up to 20.5\% for P4, and 9\% on average). With respect to Stock-Linux, LFOC reduces unfairness by 16.7\% on average.

%% file: conclusions.tex
\section{Conclusions}\label{sec:conclusions}

In this article we have presented LFOC, an OS-level cache-clustering approach that leverages cache-partitioning support in Intel-CAT enabled multicore processors to improve fairness while maintaining acceptable throughput. LFOC classifies applications into three categories according to their degree of memory-intensity and cache sensitivity, and ensures that streaming aggressor benchmarks are confined to small cache partitions so they are  isolated from cache-sensitive benchmarks, which are assigned an amount of cache space in accordance to its sensitivity. In doing so, LFOC tries to mimic the behavior of the optimal cache clustering solution, which we approximated by means of a simulator. We implemented LFOC in the Linux kernel and assessed its effectiveness on a commercial multicore platform featuring an Intel Skylake processor. Our experiments reveal that LFOC is able to deliver an average 16.7\% fairness improvement relative to stock Linux. At the same time, LFOC clearly outperforms two other existing cache-clustering approaches, one of which was specifically designed to deliver fairness~\cite{selfa-pact17}. 

Key aspects of LFOC are its lightweight clustering algorithm, the online heuristics it leverages to classify applications online, and its ability to fairly share the space on the LLC among applications by using limited monitoring information, which can be obtained at runtime without collecting performance data for every possible way count. 

%% file: main.bbl

\begin{thebibliography}{36}


\ifx \showCODEN    \undefined \def \showCODEN     #1{\unskip}     \fi
\ifx \showDOI      \undefined \def \showDOI       #1{#1}\fi
\ifx \showISBNx    \undefined \def \showISBNx     #1{\unskip}     \fi
\ifx \showISBNxiii \undefined \def \showISBNxiii  #1{\unskip}     \fi
\ifx \showISSN     \undefined \def \showISSN      #1{\unskip}     \fi
\ifx \showLCCN     \undefined \def \showLCCN      #1{\unskip}     \fi
\ifx \shownote     \undefined \def \shownote      #1{#1}          \fi
\ifx \showarticletitle \undefined \def \showarticletitle #1{#1}   \fi
\ifx \showURL      \undefined \def \showURL       {\relax}        \fi
\providecommand\bibfield[2]{#2}
\providecommand\bibinfo[2]{#2}
\providecommand\natexlab[1]{#1}
\providecommand\showeprint[2][]{arXiv:#2}

\bibitem[\protect\citeauthoryear{Brock et~al\mbox{.}}{Brock et~al\mbox{.}}{2015}]%
        {icpp-15}
\bibfield{author}{\bibinfo{person}{J. Brock} {et~al\mbox{.}}} \bibinfo{year}{2015}\natexlab{}.
\newblock \showarticletitle{Optimal Cache Partition-Sharing}. In \bibinfo{booktitle}{\emph{Proceedings of the 2015 44th International Conference on Parallel Processing (ICPP)}} \emph{(\bibinfo{series}{ICPP '15})}. \bibinfo{pages}{749--758}.
\newblock
\showISBNx{978-1-4673-7587-0}


\bibitem[\protect\citeauthoryear{Ebrahimi et~al\mbox{.}}{Ebrahimi et~al\mbox{.}}{2010}]%
        {ebrahimi10}
\bibfield{author}{\bibinfo{person}{E. Ebrahimi} {et~al\mbox{.}}} \bibinfo{year}{2010}\natexlab{}.
\newblock \showarticletitle{Fairness via source throttling: a configurable and high-performance fairness substrate for multi-core memory systems}. In \bibinfo{booktitle}{\emph{15th Int'l Conf. Architectural Support Programming Lang. and Oper. Syst. (ASPLOS 10)}}. \bibinfo{pages}{335--346}.
\newblock


\bibitem[\protect\citeauthoryear{El-Sayed et~al\mbox{.}}{El-Sayed et~al\mbox{.}}{2018a}]%
        {kpart}
\bibfield{author}{\bibinfo{person}{N. El-Sayed} {et~al\mbox{.}}} \bibinfo{year}{2018}\natexlab{a}.
\newblock \showarticletitle{KPart: A Hybrid Cache Partitioning-Sharing Technique for Commodity Multicores}. In \bibinfo{booktitle}{\emph{2018 IEEE International Symposium on High Performance Computer Architecture (HPCA)}}. \bibinfo{pages}{104--117}.
\newblock
\showISSN{2378-203X}


\bibitem[\protect\citeauthoryear{El-Sayed et~al\mbox{.}}{El-Sayed et~al\mbox{.}}{2018b}]%
        {kpart-github}
\bibfield{author}{\bibinfo{person}{N. El-Sayed} {et~al\mbox{.}}} \bibinfo{year}{2018}\natexlab{b}.
\newblock \bibinfo{title}{Source Code of KPart}.
\newblock \bibinfo{howpublished}{https://github.com/Nosayba/kpart}.
\newblock
\newblock
\shownote{Accessed: 2019-02-20.}


\bibitem[\protect\citeauthoryear{Eyerman and Eeckhout}{Eyerman and Eeckhout}{2008}]%
        {metrics-belga}
\bibfield{author}{\bibinfo{person}{S. Eyerman} {and} \bibinfo{person}{L. Eeckhout}.} \bibinfo{year}{2008}\natexlab{}.
\newblock \showarticletitle{System-Level Performance Metrics for Multiprogram Workloads}.
\newblock \bibinfo{journal}{\emph{IEEE Micro}} \bibinfo{volume}{28}, \bibinfo{number}{3} (\bibinfo{date}{May} \bibinfo{year}{2008}), \bibinfo{pages}{42--53}.
\newblock
\showISSN{0272-1732}


\bibitem[\protect\citeauthoryear{Feliu et~al\mbox{.}}{Feliu et~al\mbox{.}}{2016}]%
        {valencianos-tocs}
\bibfield{author}{\bibinfo{person}{J. Feliu} {et~al\mbox{.}}} \bibinfo{year}{2016}\natexlab{}.
\newblock \showarticletitle{Perf \& Fair: a Progress-Aware Scheduler to Enhance Performance and Fairness in {SMT} Multicores}.
\newblock \bibinfo{journal}{\emph{IEEE Trans. Comput.}} \bibinfo{volume}{PP}, \bibinfo{number}{99} (\bibinfo{year}{2016}).
\newblock
\showISSN{0018-9340}


\bibitem[\protect\citeauthoryear{Funaro, Ben-Yehuda, and Schuster}{Funaro et~al\mbox{.}}{2016}]%
        {ginseng-atc16}
\bibfield{author}{\bibinfo{person}{L. Funaro}, \bibinfo{person}{O.~A. Ben-Yehuda}, {and} \bibinfo{person}{A. Schuster}.} \bibinfo{year}{2016}\natexlab{}.
\newblock \showarticletitle{Ginseng: Market-driven LLC Allocation}. In \bibinfo{booktitle}{\emph{Proceedings of the 2016 USENIX Annual Technical Conference}} \emph{(\bibinfo{series}{USENIX ATC '16})}. \bibinfo{pages}{295--308}.
\newblock
\showISBNx{978-1-931971-30-0}


\bibitem[\protect\citeauthoryear{Garcia-Garcia, Casas, and Saez}{Garcia-Garcia et~al\mbox{.}}{2019}]%
        {pbbcache-github}
\bibfield{author}{\bibinfo{person}{A. Garcia-Garcia}, \bibinfo{person}{J. Casas}, {and} \bibinfo{person}{J.~C. Saez}.} \bibinfo{year}{2019}\natexlab{}.
\newblock \bibinfo{title}{{P}{B}{B}{C}ache: A parallel branch-and-bound based cache-partitioning simulator}.
\newblock \bibinfo{howpublished}{https://github.com/pbbcache/cachesim}.
\newblock
\newblock
\shownote{Accessed: 2019-05-10.}


\bibitem[\protect\citeauthoryear{Garcia-Garcia, Saez, and Prieto-Matias}{Garcia-Garcia et~al\mbox{.}}{2018}]%
        {camps}
\bibfield{author}{\bibinfo{person}{A. Garcia-Garcia}, \bibinfo{person}{J.~C. Saez}, {and} \bibinfo{person}{M. Prieto-Matias}.} \bibinfo{year}{2018}\natexlab{}.
\newblock \showarticletitle{Contention-Aware Fair Scheduling for Asymmetric Single-{ISA} Multicore Systems}.
\newblock \bibinfo{journal}{\emph{IEEE Trans. Comput.}} \bibinfo{volume}{67}, \bibinfo{number}{12} (\bibinfo{date}{Dec} \bibinfo{year}{2018}), \bibinfo{pages}{1703--1719}.
\newblock
\showISSN{0018-9340}


\bibitem[\protect\citeauthoryear{Khan et~al\mbox{.}}{Khan et~al\mbox{.}}{2014}]%
        {khan-hpca14}
\bibfield{author}{\bibinfo{person}{S.~M. Khan} {et~al\mbox{.}}} \bibinfo{year}{2014}\natexlab{}.
\newblock \showarticletitle{Improving cache performance using read-write partitioning}. In \bibinfo{booktitle}{\emph{20th {IEEE} International Symposium on High Performance Computer Architecture, {HPCA} 2014}}. \bibinfo{pages}{452--463}.
\newblock


\bibitem[\protect\citeauthoryear{Lo et~al\mbox{.}}{Lo et~al\mbox{.}}{2015}]%
        {heracles15}
\bibfield{author}{\bibinfo{person}{D. Lo} {et~al\mbox{.}}} \bibinfo{year}{2015}\natexlab{}.
\newblock \showarticletitle{Heracles: improving resource efficiency at scale}. In \bibinfo{booktitle}{\emph{Proc. of the 42nd Annual International Symposium on Computer Architecture}}. \bibinfo{pages}{450--462}.
\newblock


\bibitem[\protect\citeauthoryear{Love}{Love}{2010}]%
        {love-lkd}
\bibfield{author}{\bibinfo{person}{R. Love}.} \bibinfo{year}{2010}\natexlab{}.
\newblock \bibinfo{booktitle}{\emph{Linux Kernel Development} (\bibinfo{edition}{3rd} ed.)}.
\newblock \bibinfo{publisher}{Addison-Wesley Professional}.
\newblock
\showISBNx{0672329468, 9780672329463}


\bibitem[\protect\citeauthoryear{Manikantan, Rajan, and Govindarajan}{Manikantan et~al\mbox{.}}{2012}]%
        {prism-isca12}
\bibfield{author}{\bibinfo{person}{R. Manikantan}, \bibinfo{person}{K. Rajan}, {and} \bibinfo{person}{R. Govindarajan}.} \bibinfo{year}{2012}\natexlab{}.
\newblock \showarticletitle{Probabilistic Shared Cache Management (PriSM)}. In \bibinfo{booktitle}{\emph{Proceedings of the 39th Annual International Symposium on Computer Architecture}} \emph{(\bibinfo{series}{ISCA '12})}. \bibinfo{pages}{428--439}.
\newblock
\showISBNx{978-1-4503-1642-2}


\bibitem[\protect\citeauthoryear{Mittal}{Mittal}{2017}]%
        {survey-cachepart}
\bibfield{author}{\bibinfo{person}{S. Mittal}.} \bibinfo{year}{2017}\natexlab{}.
\newblock \showarticletitle{A Survey of Techniques for Cache Partitioning in Multicore Processors}.
\newblock \bibinfo{journal}{\emph{ACM Comput. Surv.}} \bibinfo{volume}{50}, \bibinfo{number}{2}, Article \bibinfo{articleno}{27} (\bibinfo{date}{May} \bibinfo{year}{2017}), \bibinfo{numpages}{27:1--27:39}~pages.
\newblock
\showISSN{0360-0300}


\bibitem[\protect\citeauthoryear{Morad et~al\mbox{.}}{Morad et~al\mbox{.}}{2016}]%
        {morad-jpdc16}
\bibfield{author}{\bibinfo{person}{T.Y. Morad} {et~al\mbox{.}}} \bibinfo{year}{2016}\natexlab{}.
\newblock \showarticletitle{EFS: Energy-Friendly Scheduler for memory bandwidth constrained systems}.
\newblock \bibinfo{journal}{\emph{J. Parallel and Distrib. Comput.}}  \bibinfo{volume}{95} (\bibinfo{year}{2016}), \bibinfo{pages}{3 -- 14}.
\newblock
\showISSN{0743-7315}


\bibitem[\protect\citeauthoryear{Mukkara, Beckmann, and Sanchez}{Mukkara et~al\mbox{.}}{2016}]%
        {whirlpool}
\bibfield{author}{\bibinfo{person}{A. Mukkara}, \bibinfo{person}{N. Beckmann}, {and} \bibinfo{person}{D. Sanchez}.} \bibinfo{year}{2016}\natexlab{}.
\newblock \showarticletitle{Whirlpool: Improving Dynamic Cache Management with Static Data Classification}. In \bibinfo{booktitle}{\emph{Proc. of the 21st Int'l Conf. on Arch. Support for Programming Lang. and Oper. Syst.}} \emph{(\bibinfo{series}{ASPLOS '16})}. \bibinfo{pages}{113--127}.
\newblock
\showISBNx{978-1-4503-4091-5}


\bibitem[\protect\citeauthoryear{Mutlu and Moscibroda}{Mutlu and Moscibroda}{2007}]%
        {stall-time-dram}
\bibfield{author}{\bibinfo{person}{O. Mutlu} {and} \bibinfo{person}{T. Moscibroda}.} \bibinfo{year}{2007}\natexlab{}.
\newblock \showarticletitle{Stall-Time Fair Memory Access Scheduling for Chip Multiprocessors}. In \bibinfo{booktitle}{\emph{40th Ann. IEEE/ACM Int'l Symp. on Microarchitecture (MICRO 07)}}. \bibinfo{pages}{146--160}.
\newblock
\showISBNx{0-7695-3047-8}


\bibitem[\protect\citeauthoryear{Nguyen}{Nguyen}{2016}]%
        {cat}
\bibfield{author}{\bibinfo{person}{K. Nguyen}.} \bibinfo{year}{2016}\natexlab{}.
\newblock \bibinfo{title}{Introduction to Cache Allocation Technology in the Intel Xeon Processor E5 v4 Family}.
\newblock \bibinfo{howpublished}{\url{https://software.intel.com/en-us/articles/introduction-to-cache-allocation-technology}}.
\newblock
\newblock
\shownote{Accessed: 2019-03-20.}


\bibitem[\protect\citeauthoryear{Qureshi and Patt}{Qureshi and Patt}{2006}]%
        {ucp}
\bibfield{author}{\bibinfo{person}{M.K. Qureshi} {and} \bibinfo{person}{Y.N. Patt}.} \bibinfo{year}{2006}\natexlab{}.
\newblock \showarticletitle{Utility-Based Cache Partitioning: A Low-Overhead, High-Performance, Runtime Mechanism to Partition Shared Caches}. In \bibinfo{booktitle}{\emph{Proceedings of MICRO 06}}. \bibinfo{pages}{423--432}.
\newblock


\bibitem[\protect\citeauthoryear{Saez et~al\mbox{.}}{Saez et~al\mbox{.}}{2017a}]%
        {pmctrack-compj}
\bibfield{author}{\bibinfo{person}{J.C. Saez} {et~al\mbox{.}}} \bibinfo{year}{2017}\natexlab{a}.
\newblock \showarticletitle{{PMCTrack}: Delivering Performance Monitoring Counter Support to the {OS} Scheduler}.
\newblock \bibinfo{journal}{\emph{Comput. J.}} \bibinfo{volume}{60}, \bibinfo{number}{1} (\bibinfo{year}{2017}), \bibinfo{pages}{60--85}.
\newblock


\bibitem[\protect\citeauthoryear{Saez et~al\mbox{.}}{Saez et~al\mbox{.}}{2017b}]%
        {acfs-jpdc}
\bibfield{author}{\bibinfo{person}{J.C. Saez} {et~al\mbox{.}}} \bibinfo{year}{2017}\natexlab{b}.
\newblock \showarticletitle{Towards completely fair scheduling on asymmetric single-{ISA} multicore processors}.
\newblock \bibinfo{journal}{\emph{J. Parallel and Distrib. Comput.}}  \bibinfo{volume}{102} (\bibinfo{year}{2017}), \bibinfo{pages}{115 -- 131}.
\newblock
\showISSN{0743-7315}


\bibitem[\protect\citeauthoryear{Saez, Gomez, and Prieto}{Saez et~al\mbox{.}}{2008}]%
        {nwc-sched}
\bibfield{author}{\bibinfo{person}{J.C. Saez}, \bibinfo{person}{J.I. Gomez}, {and} \bibinfo{person}{M. Prieto}.} \bibinfo{year}{2008}\natexlab{}.
\newblock \showarticletitle{Improving Priority Enforcement via Non-Work-Conserving Scheduling}. In \bibinfo{booktitle}{\emph{ICPP '08: Proceedings of the 2008 37th International Conference on Parallel Processing}}. \bibinfo{pages}{99--106}.
\newblock
\showISBNx{978-0-7695-3374-2}


\bibitem[\protect\citeauthoryear{Scolari, Bartolini, and Santambrogio}{Scolari et~al\mbox{.}}{2016}]%
        {taco16}
\bibfield{author}{\bibinfo{person}{A. Scolari}, \bibinfo{person}{D.B. Bartolini}, {and} \bibinfo{person}{M.D. Santambrogio}.} \bibinfo{year}{2016}\natexlab{}.
\newblock \showarticletitle{A Software Cache Partitioning System for Hash-Based Caches}.
\newblock \bibinfo{journal}{\emph{ACM Trans. Archit. Code Optim.}} \bibinfo{volume}{13}, \bibinfo{number}{4}, Article \bibinfo{articleno}{57} (\bibinfo{date}{Dec.} \bibinfo{year}{2016}), \bibinfo{numpages}{57:1--57:24}~pages.
\newblock
\showISSN{1544-3566}


\bibitem[\protect\citeauthoryear{Selfa et~al\mbox{.}}{Selfa et~al\mbox{.}}{2017}]%
        {selfa-pact17}
\bibfield{author}{\bibinfo{person}{V. Selfa} {et~al\mbox{.}}} \bibinfo{year}{2017}\natexlab{}.
\newblock \showarticletitle{Application Clustering Policies to Address System Fairness with Intel's Cache Allocation Technology}. In \bibinfo{booktitle}{\emph{2017 26th International Conference on Parallel Architectures and Compilation Techniques (PACT)}}. \bibinfo{pages}{194--205}.
\newblock


\bibitem[\protect\citeauthoryear{Sherwood, Calder, and Emer}{Sherwood et~al\mbox{.}}{1999}]%
        {ics-99}
\bibfield{author}{\bibinfo{person}{T. Sherwood}, \bibinfo{person}{B. Calder}, {and} \bibinfo{person}{J. Emer}.} \bibinfo{year}{1999}\natexlab{}.
\newblock \showarticletitle{Reducing Cache Misses Using Hardware and Software Page Placement}. In \bibinfo{booktitle}{\emph{Proceedings of the 13th International Conference on Supercomputing}} \emph{(\bibinfo{series}{ICS '99})}. \bibinfo{pages}{155--164}.
\newblock
\showISBNx{1-58113-164-X}


\bibitem[\protect\citeauthoryear{Subramanian et~al\mbox{.}}{Subramanian et~al\mbox{.}}{2015}]%
        {subramanian-micro15}
\bibfield{author}{\bibinfo{person}{L. Subramanian} {et~al\mbox{.}}} \bibinfo{year}{2015}\natexlab{}.
\newblock \showarticletitle{The Application Slowdown Model: Quantifying and Controlling the Impact of Inter-application Interference at Shared Caches and Main Memory}. In \bibinfo{booktitle}{\emph{Proceedings of the 48th International Symposium on Microarchitecture}} \emph{(\bibinfo{series}{MICRO-48})}. \bibinfo{pages}{62--75}.
\newblock
\showISBNx{978-1-4503-4034-2}


\bibitem[\protect\citeauthoryear{Van~Craeynest et~al\mbox{.}}{Van~Craeynest et~al\mbox{.}}{2013}]%
        {eqp-belga}
\bibfield{author}{\bibinfo{person}{K. Van~Craeynest} {et~al\mbox{.}}} \bibinfo{year}{2013}\natexlab{}.
\newblock \showarticletitle{Fairness-aware scheduling on single-{ISA} heterogeneous multi-cores}. In \bibinfo{booktitle}{\emph{22nd Int'l Conf. Parallel Arch. Compilation Techniques (PACT 13)}}. \bibinfo{pages}{177--187}.
\newblock


\bibitem[\protect\citeauthoryear{Wang and Chen}{Wang and Chen}{2014}]%
        {wang-micro14}
\bibfield{author}{\bibinfo{person}{R. Wang} {and} \bibinfo{person}{L. Chen}.} \bibinfo{year}{2014}\natexlab{}.
\newblock \showarticletitle{Futility Scaling: High-Associativity Cache Partitioning}. In \bibinfo{booktitle}{\emph{Proceedings of the 47th Annual IEEE/ACM International Symposium on Microarchitecture}} \emph{(\bibinfo{series}{MICRO-47})}. \bibinfo{pages}{356--367}.
\newblock
\showISBNx{978-1-4799-6998-2}


\bibitem[\protect\citeauthoryear{Xu et~al\mbox{.}}{Xu et~al\mbox{.}}{2012}]%
        {xu-sigmetrics12}
\bibfield{author}{\bibinfo{person}{D. Xu} {et~al\mbox{.}}} \bibinfo{year}{2012}\natexlab{}.
\newblock \showarticletitle{Providing Fairness on Shared-memory Multiprocessors via Process Scheduling}. In \bibinfo{booktitle}{\emph{Proc. ACM Int'l Conf. Measurement and Modeling Comp. Syst. (SIGMETRICS 12)}}. \bibinfo{pages}{295--306}.
\newblock
\showISBNx{978-1-4503-1097-0}


\bibitem[\protect\citeauthoryear{Ye et~al\mbox{.}}{Ye et~al\mbox{.}}{2014}]%
        {pact-14}
\bibfield{author}{\bibinfo{person}{Y. Ye} {et~al\mbox{.}}} \bibinfo{year}{2014}\natexlab{}.
\newblock \showarticletitle{COLORIS: A Dynamic Cache Partitioning System Using Page Coloring}. In \bibinfo{booktitle}{\emph{Proceedings of the 23rd International Conference on Parallel Architectures and Compilation}} \emph{(\bibinfo{series}{PACT '14})}. \bibinfo{pages}{381--392}.
\newblock
\showISBNx{978-1-4503-2809-8}


\bibitem[\protect\citeauthoryear{Yu and Petrov}{Yu and Petrov}{2010}]%
        {yu-petrof}
\bibfield{author}{\bibinfo{person}{C. Yu} {and} \bibinfo{person}{P. Petrov}.} \bibinfo{year}{2010}\natexlab{}.
\newblock \showarticletitle{Off-chip Memory Bandwidth Minimization Through Cache Partitioning for Multi-core Platforms}. In \bibinfo{booktitle}{\emph{Proceedings of the 47th Design Automation Conference}} \emph{(\bibinfo{series}{DAC '10})}. \bibinfo{pages}{132--137}.
\newblock
\showISBNx{978-1-4503-0002-5}


\bibitem[\protect\citeauthoryear{Yun et~al\mbox{.}}{Yun et~al\mbox{.}}{2014}]%
        {palloc}
\bibfield{author}{\bibinfo{person}{H. Yun} {et~al\mbox{.}}} \bibinfo{year}{2014}\natexlab{}.
\newblock \showarticletitle{{PALLOC}: {DRAM} bank-aware memory allocator for performance isolation on multicore platforms}. In \bibinfo{booktitle}{\emph{20th Real-Time Embedded Tech. and Applications Symp. (RTAS 14)}}. \bibinfo{pages}{155--166}.
\newblock


\bibitem[\protect\citeauthoryear{Yun et~al\mbox{.}}{Yun et~al\mbox{.}}{2016}]%
        {heechul-transc16}
\bibfield{author}{\bibinfo{person}{H. Yun} {et~al\mbox{.}}} \bibinfo{year}{2016}\natexlab{}.
\newblock \showarticletitle{Memory Bandwidth Management for Efficient Performance Isolation in Multi-Core Platforms}.
\newblock \bibinfo{journal}{\emph{IEEE Trans. Comput.}} \bibinfo{volume}{65}, \bibinfo{number}{2} (\bibinfo{date}{Feb} \bibinfo{year}{2016}), \bibinfo{pages}{562--576}.
\newblock
\showISSN{0018-9340}


\bibitem[\protect\citeauthoryear{Zhang, Dwarkadas, and Shen}{Zhang et~al\mbox{.}}{2009}]%
        {xiao-eurosys09}
\bibfield{author}{\bibinfo{person}{X. Zhang}, \bibinfo{person}{S. Dwarkadas}, {and} \bibinfo{person}{K. Shen}.} \bibinfo{year}{2009}\natexlab{}.
\newblock \showarticletitle{Towards Practical Page Coloring-based Multicore Cache Management}. In \bibinfo{booktitle}{\emph{Proceedings of the 4th ACM European Conference on Computer Systems}} \emph{(\bibinfo{series}{EuroSys '09})}. \bibinfo{pages}{89--102}.
\newblock
\showISBNx{978-1-60558-482-9}


\bibitem[\protect\citeauthoryear{Zhu and Erez}{Zhu and Erez}{2016}]%
        {dirigent-asplos16}
\bibfield{author}{\bibinfo{person}{H. Zhu} {and} \bibinfo{person}{M. Erez}.} \bibinfo{year}{2016}\natexlab{}.
\newblock \showarticletitle{Dirigent: Enforcing QoS for Latency-Critical Tasks on Shared Multicore Systems}. In \bibinfo{booktitle}{\emph{Proc. of the 21st Int'l Conf. on Architectural Support for Programming Languages and Operating Systems}} \emph{(\bibinfo{series}{ASPLOS '16})}. \bibinfo{pages}{33--47}.
\newblock
\showISBNx{978-1-4503-4091-5}


\bibitem[\protect\citeauthoryear{Zhuravlev et~al\mbox{.}}{Zhuravlev et~al\mbox{.}}{2012}]%
        {survey-contention}
\bibfield{author}{\bibinfo{person}{S. Zhuravlev} {et~al\mbox{.}}} \bibinfo{year}{2012}\natexlab{}.
\newblock \showarticletitle{Survey of Scheduling Techniques for Addressing Shared Resources in Multicore Processors}.
\newblock \bibinfo{journal}{\emph{ACM Comput. Surv.}} \bibinfo{volume}{45}, \bibinfo{number}{1}, Article \bibinfo{articleno}{4} (\bibinfo{date}{Dec.} \bibinfo{year}{2012}), \bibinfo{numpages}{28}~pages.
\newblock
\showISSN{0360-0300}


\end{thebibliography}
